\documentclass[proof]{WileyASNA-v1}
\usepackage{graphicx}
\usepackage{amsmath}
\articletype{Article Type}%

\received{TBD}
\revised{TBD}
\accepted{TBD}
\pdfoutput=1
\begin{document}

\title{Detecting Radio Frequency Interference in radio-antenna arrays 
with the Recurrent Neural Network algorithm}

\author[1]{P.R. Burd$^{1}$}

\author[1]{K. Mannheim$^{1}$}

\author[1]{T. M\"arz$^{1}$}

\author[1]{J. Ringholz$^{1}$}

\author[1]{A. Kappes$^{1}$}

\author[1]{M. Kadler$^{1}$}

\authormark{P.R. Burd {et al}}

\address[1]{\orgdiv{ITPA W\"urzburg}, \orgname{ITPA}, \orgaddress{\state{Bavaria}, \country{Germany}}}

\corres{P.R. Burd \email{paul.r.burd@astro.uni-wuerzburg.de}}

\presentaddress{Emil-Fischer-Strasse 31, 97074 W\"urzburg, Germany}

\abstract{Signal artefacts due to Radio Frequency Interference (RFI) are a common nuisance in radio astronomy.  Conventionally,
the RFI-affected data are tagged by an expert data analyst in order
to warrant data quality.  In view of the increasing data rates obtained
with interferometric radio telescope arrays, automatic data filtering procedures are mandatory.
Here, we present results from the implementation of a RFI-detecting recurrent neural network (RNN) employing long-short term memory (LSTM) cells. 
For the training of the algorithm, a discrete model was used that distinguishes RFI and non-RFI data, respectively, based on the amplitude 
information from radio interferometric observations with the GMRT at $610\, \mathrm{MHz}$. 
The performance of the RNN is evaluated by analyzing a confusion matrix. The true positive and true negative rates of the network are $\approx 99.9\,\%$ and $\approx 97.9\,\%$, respectively. However, the overall efficiency of the network is $\approx 30\%$ due to the fact that a large amount non-RFI data are classified as being contaminated by RFI. Matthews correlation coefficient is ~0.42 suggesting that a still more refined training model is required.}

\keywords{methods: data analysis -- methods: numerical -- radio continuum: general -- radio lines: general -- techniques: interferometric}

\jnlcitation{\cname{%
\author{Burd P.R.}, 
\author{K. Mannheim},
\author{T. M\"arz}, 
\author{J. Ringholz},
\author{A. Kappes}, and 
\author{M. Kadler}} (\cyear{2018}), 
\ctitle{RFI detection with a Recurrent Neural Network}, \cjournal{Q.J.R. Meteorol. Soc.}, \cvol{TBA}.}

\maketitle

\section{RFI mitigation and the machine-learning approach}
Radio Frequency Interference (RFI) collectively denominates artefacts in the data of radio telescopes due to GPS transmitters, 
cell-phones, micowave ovens, pasture fences, power supply lines, thunderstorms, or similar radio emitters in the vicinity of the telescopes
with their high-sensitivity receivers.  RFI can spoil the data quality and impede calibration efforts, or mimic false astrophysical sources.
It is therefore imperative to filter those signals before the calibration and imaging analysis can proceed, see \citep{Fridman2001} and \,\citep{Offringa2010_1}. The so-called ``SumThreshold'' is a threshold-based and widely used hard-coded algorithm mitigating RFI, see \,\citep{Offringa2010_1},\,\citep{Offringa2010_2}, and \,\citep{Peck2013}.\newline
Due to the random nature and diversity of the RFI signal shapes in the spatial and frequency domains, applying a
fixed set of rules and cuts in data space generally do not suffice to eliminate RFI.  Instead, the time-consuming effort of an expert data analyst
is conventionally involved to deal with the observed complexity.
Machine-learning (ML) algorithm may be superior in providing the required flexibility and efficiency.
As a matter of fact, \citep{Akeret2017} and  \citep{Czech2018} have recently successfully applied different models of deep neural networks (DNNs) to identify RFI in data from single-dish radio telescopes.\newline
In this paper, we employ the Recurrent Neural Network (RNN) algorithm for RFI detection in data from an interferometric array of radio telescopes.
The RNN makes best use of data where any kind of order is relevant, when equipped with a long short term memory (LSTM) cell, cf. \citep{Hochreiter1997}. In this context the mentioned order can be a frequency (channel)-order, time-order or baseline-order. For the training of the algorithm, we used data obtained with the Giant Metre-Wave Radio Telescope (GMRT), see \citep{Ananthakrishnan1995} which is heavily polluted by RFI. In Sec.\ref{sec2} the data processing and the training model are described and discussed. Sec.\ref{sec3} briefly describes  the RNN architecture. The performance is discussed in Sec.\ref{sec4} with respect to the implications on the chosen RNN architecture and data modeling. 
\section{ Data Processing and Training Model}
\label{sec2}
The GMRT data, recorded at $610\, \mathrm{MHz}$, with a bandwidth of $33\, \mathrm{kHz}$ and devided into 256 channels, are available in the GMRT data archive\footnote{\url{https://naps.ncra.tifr.res.in/goa/mt/search/basicSearch}} under the project code $28\_ 029$ and the observation numbers $7779$ and $7788$. The data from the GMRT data archive are provided in FITS format. The GMRT consits of 30 antennas, thus leading to 435 $\left(\frac{N(N-1)}{2}\right)$ baselines, being able to crate 435 visibilities at any given time step and channel. 
Using CASA's python application programming interface (API), casacore, see \citep{McMullin2007}, data blocks with the dimensions 
$\underbrace{[\text{TIMESTEPS }, \text{POLARIZATION},\text{BASELINE},\text{CHANNELS}]}_{\text{TS, POL, BL, CHAN}}$ containing the amplitude information of the observations, are created.
At this point, $95\%$ of the number of data blocks, with respect to the time step, are randomly selected and used to train the RNN. The remaining $5\%$ of the data are used to test the performance of the RNN.\newline
It is worth mentioning, that also the information of the phase, derived from the visibility, and the differences of phase (amplitudes) with respect to the channel, baseline and/or time order can be used to find RFI and to train the RNN. However this study focuses only on the amplitudes as a first step to assess the method's potential on this level.\newline
We train the RNN to be sensitive to the sequence of data with respect to the channels, meaning the training block has the form 
$$[\underbrace{(\text{TS}\times \text{POL}, \text{BL}), (\text{TS}\times \text{POL}, \text{BL})...}_{\text{CHAN} \text{ times}}].$$ 
This means for this first approach we feed each time step, baseline and polarization per channel into the RNN. This becomes important when interpreting the resulting classifications of the RNN. 
Table \ref{tab2} lists all dimensions of the axis in the data block. The training and test data set contain the same dimensions along the channel (CHAN) and baseline (BL) axis, the time step times polarization axis however, as mentioned above, is split into $95\%$ and $5\%$ of the total available dimension along this axis, respectively.
\begin{center}
\begin{table}[t]%
\centering
\caption{The dimensions of each axis in the data block are listed here. The training and the test data set have the same dimensions along the channel (CHAN) and baseline (BL) axis, however the timestep times polarization is split into $95\%$ and $5\%$ of the entire data set, respectively.\label{tab2}}%
\tabcolsep=0pt%
\begin{tabular*}{20pc}{@{\extracolsep\fill}lcc@{\extracolsep\fill}}
\toprule
&\textbf{training data} & \textbf{test data}  \\
\midrule
TS $\times$ POL& 3395& 179\\
CHAN & 256 & 256\\
BL & 435&435\\
\bottomrule
\end{tabular*}
\end{table}
\end{center}
Before feeding the data into the RNN, the amplitudes are re-scaled between zero and one. This procedure results in a number of data blocks which corresponds to the number of time steps multiplied by the polarization where each amplitude per channel and baseline can be fed into the RNN.\newline
To train the network, a simple model is built to label certain channels as RFI contaminated. The algorithm scans the amplitudes in each channel. Within the channel interval $i-3 < i < i+3$, the median is calculated. If in one channel the amplitude value is larger than five times the median within the neighboring range, the channel is labeled as being RFI contaminated. In this way, an array with zeros and ones is created where a zero-label denotes the RFI-free channel and a one-label denotes the RFI contaminated channel. This model is trained to the RNN to find RFIs in certain channels. 

\section{RNN architecture}
\label{sec3}
\begin{figure}
\includegraphics[width=0.5\textwidth]{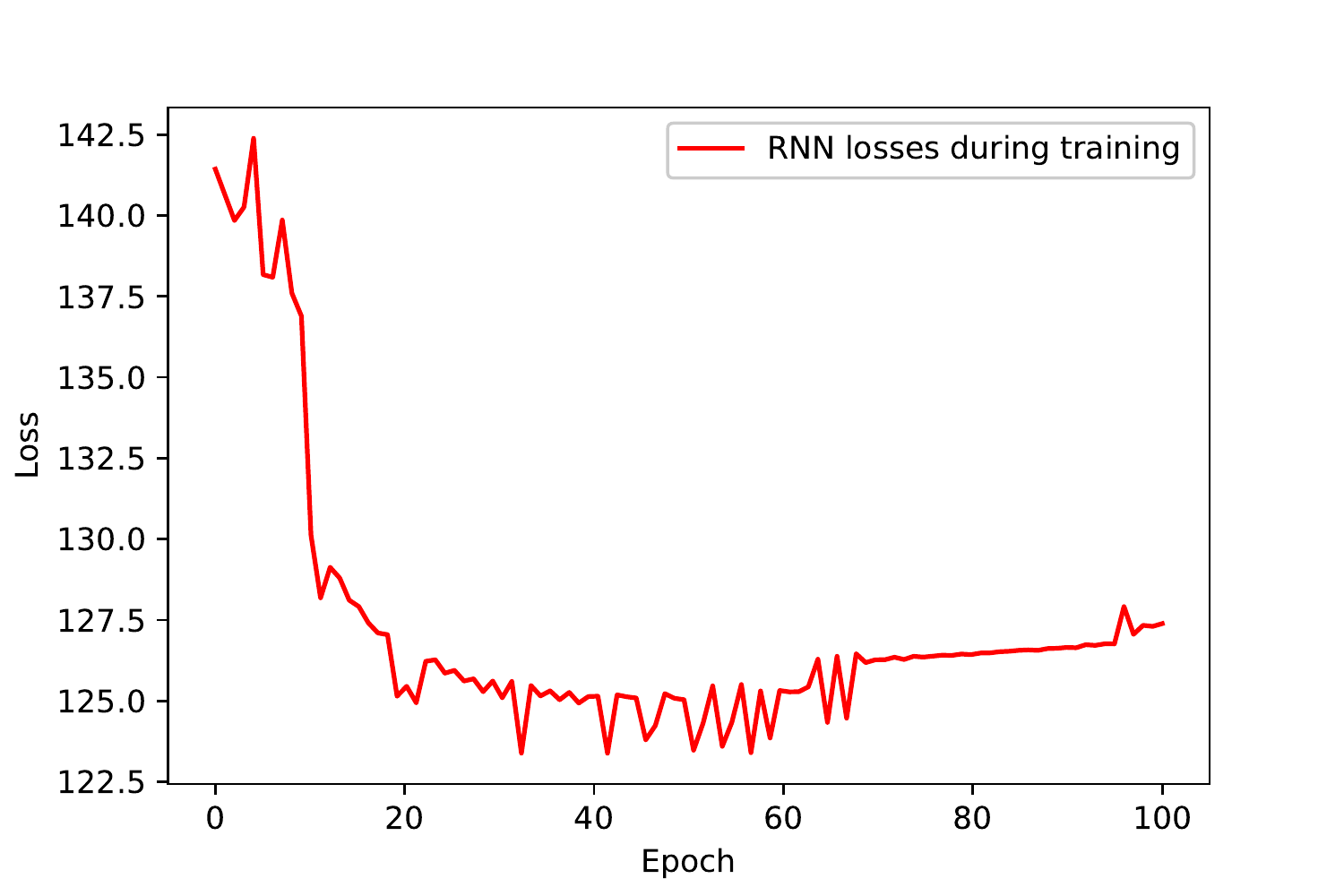}
\caption{Plot of loss function versus training epoch number. After the initially rapid drop of the losses after about 20 epochs,
the minimum of the loss function is readily achieved between 30 and 60 iterations indicating successfull training. \label{fig2}}
\end{figure}

The network is coded using the software package TensorFlow-GPU1.4.0 (TF), see \citep{TF}. The implementation addresses the CUDA cores on two GeForce GTX1080 Ti boards, which were used to train the RNN with the CUDA8.0 version. We utilize tensorflow's LSTM cell as described by \citep{Hochreiter1997} to implement the RNN. The RNN as a whole consists of 1024 of such LSTM cells. The sigmoid function is used as the activation function within each LSTM cell. 
To measure how well the RNN's model fits the data, we deploy the TFs  $\text{sigmoid}\_\text{cross}\_ \text{entropy} \_ \text{with} \_ \text{logits}$ as cost-function which is minimized using the Adam optimizer \citep{Kingma2014}. Figure.\ref{fig2} illustrates the losses during the training process.  The Adam optimizer tries to find a global minimum for the cost function. As can be seen in Fig.\ref{fig2}, a shallow minimum is found between 40 and 60 epochs. However, the result must be handled with some
caution. Due to the non-linearity of the problem, the global minimum might still be outside of the range of the numerical accuracy reached after
100 iterations where we stopped for practical reasons.

\section{Performance}
\label{sec4}
\label{perf}
\begin{figure}
\includegraphics[width=0.5\textwidth]{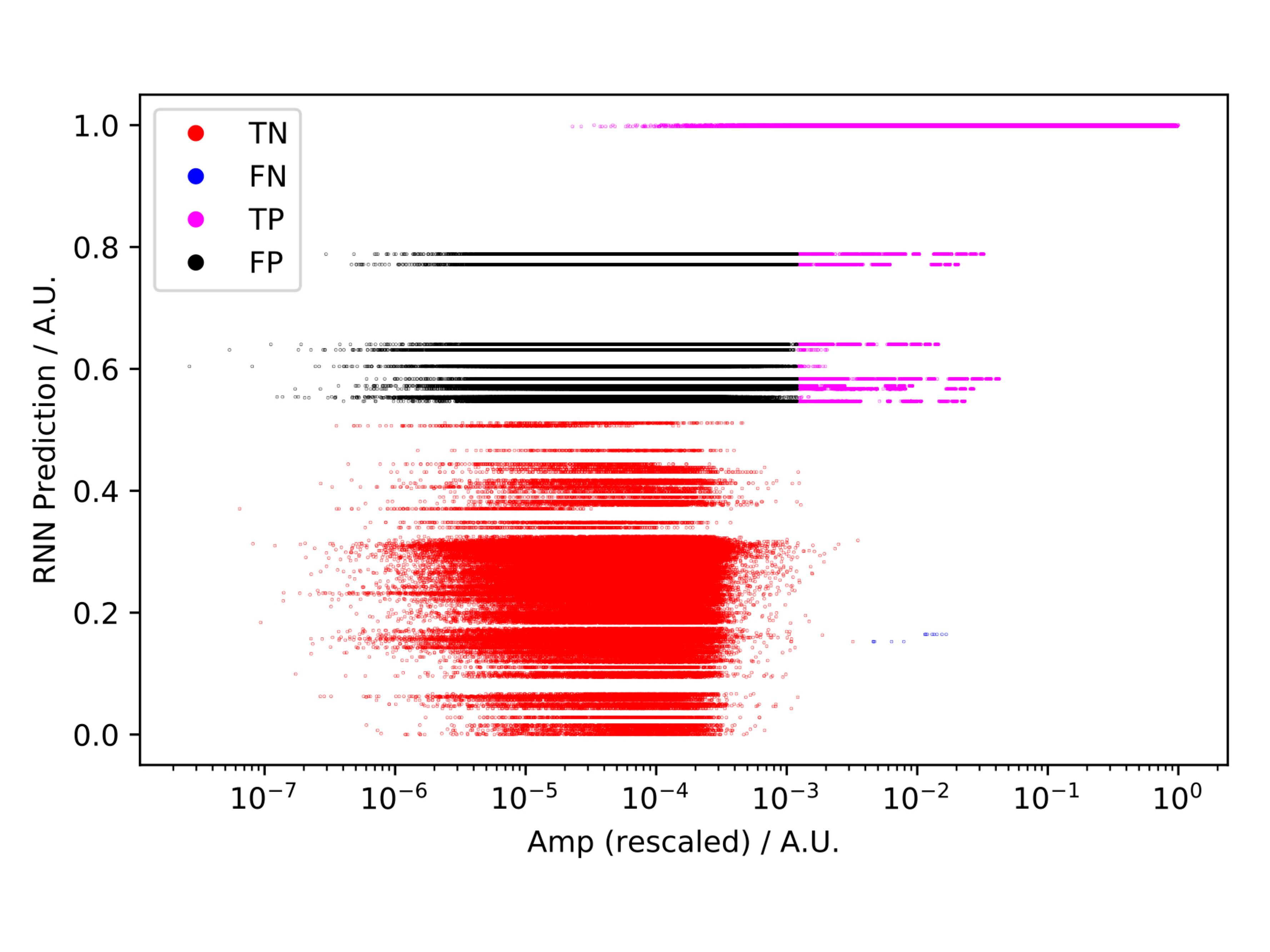}
\caption{The RNN classification is rescaled between zero and one and plotted against the rescaled amplitude.The colors indicate the different categories of detection.  The data points colored in magenta show the TP detection. Red-colored datapoints depict the TN detection. The FN and FP detections are illustrated in blue and black, respectively. Note: the data density is thinned out in the TN region (red) by a factor of 100 and in the FN rgion (black) by a factor of 10 to reduce the size of the file while preserving the information for visual inspection.\label{fig1}}
\end{figure}

The result of the RFI classification capability of the RNN is illustrated in Fig.\ref{fig1}. The different sections in the plot can be interpreted as follows:
\begin{itemize}
\item true positive (TP) classifications (magenta): correctly classified RFI
\item true negative (TN) classifications (red): correctly classified non-RFI
\item false negative (FN) classifications (blue): incorrectly classified RFI
\item false positive (FP) classifications (black): incorrectly classified non-RFI 
\end{itemize}
The amount of data points in each category is summarized in the confusion matrix:
\begin{equation}
\label{confmatr}
\text{confusion matrix} = 
\begin{pmatrix} \text{TP} & \text{FP}\\ \text{FN} & \text{TN} \end{pmatrix} = \begin{pmatrix} 91161 & 415179\\ 128 & 19426972 \end{pmatrix} 
\end{equation}
The confusion matrix is evaluated according to \citep{Boughorbel2017}, \citep{Fawcett2006} and \citep{Powers2011}. The results are summarized in Tab.\ref{tab1}. In the following, we discuss the RNN efficiency for RFI detection:

\paragraph{\it Accuracy} The accuracy with which the network separates RFI and non-RFI signals according to following Eq.\ref{eq_acc}
\begin{equation}
\label{eq_acc}
\text{ accuracy} = \frac{\text{TP}+\text{TN}}{\text{TP}+\text{TN}+\text{FP}+\text{FN}},
\end{equation}

 amounts to $\approx 98\,\%$. The high value
reflects the fact that most data points are within the TN (see Fig.\ref{fig1} category as seen in Eq.\ref{confmatr}). 
However, the value alone is not sufficient to assess the full performance.

\paragraph{\it Positive predictive value and false discovery rate}
The positive predictive value (PPV) is $\approx 18\,\%$. The PPV is defined as the fraction of the data correctly identified by the network as RFI 
compared to all data, including incorrectly identified non-RFI data (black and magenta illustrated data points in Fig.\ref{fig1}), see Eq.\ref{eq_PPV}.
\begin{equation}
\label{eq_PPV}
\text{ PPV} = \frac{\text{TP}}{\text{TP} + \text{FP}}
\end{equation}
The relatively low value for the PPV and the relatively high value for the false discovery rate FDR ($\approx 82\,\%$), which is the rate with which non-RFI signals are classified as RFI, see Eq.\ref{eq_FDR}
\begin{equation}
\label{eq_FDR}
\text{ FDR} = \frac{\text{FP}}{\text{FP} + \text{TP}},
\end{equation}
is due to the fact that for each channel all baselines, time steps and polarizations are considered. If any of those are classified to be RFI polluted, the entire channel is flagged as such. The method can obviously be further refined to improve its overall efficiency by employing a less
simplistic approach.

 \paragraph{\it Negative predictive value and false omission rate} 
The negative predictive value (NPV) is $\approx 99\,\%$. The NPV states which data points are correctly classified as non-RFI compared to all data points which are identified as non-RFI (red and blue marked data points in Fig.\ref{fig1}), see Eq.\ref{eq_NPV}.
\begin{equation}
\label{eq_NPV}
\text{ NPV} = \frac{\text{TN}}{\text{TN} + \text{FN}}
\end{equation}
Due to the fact that most data points are in the TN category ($\approx 19\times 10^6$) compared to 128 in the FN category it becomes clear the the NPV converges to one while the false omission rate (FOR), Eq.\ref{eq_FOR}, which is the rate with which RFI signals are not identified as such, goes to zero.
\begin{equation}
\label{eq_FOR}
\text{ FOR} = \frac{\text{FN}}{\text{FN}+ \text{TN}}
\end{equation}

\paragraph{\it True positive and false negative rate} 
The true positive rate, Eq.\ref{eq_TPR}, is $\approx 99.86\,\%$, due to the TP being two orders of magnitude larger than the FN value. It describes the network ability to successfully predict RFI.
\begin{equation}
\label{eq_TPR}
\text{ TPR} =\frac{\text{TP}}{\text{TP} + \text{FN}} 
\end{equation}
The false negative rate (FNR), see Eq.\ref{eq_FNR} on the other hand is $\approx 0.14\,\%$.
\begin{equation}
\label{eq_FNR}
\text{ FNR} = \frac{\text{FN}}{\text{FN} + \text{TP}}
\end{equation}

\paragraph{\it True negative rate and false positive rate}
The true negative rate (TNR), Eq.\ref{eq_TNR}, is $\approx 97.91\,\%$. Similar as in the paragraph before the TN is two orders of magnitude larger than the FP value.
\begin{equation}
\label{eq_TNR}
\text{ TNR} =\frac{\text{TN}}{\text{TN} + \text{FP}}
\end{equation}
The false positive rate (FPR), see Eq.\ref{eq_FPR}, is $\approx 2.09\,\%$.
\begin{equation}
\label{eq_FPR}
\text{ FPR} = \frac{\text{FP}}{\text{FP}+\text{TN}}
\end{equation}

\paragraph{\it Matthews correlation coefficient and F1 score}
The Matthews correlation coefficient (MCC), see Eq.\ref{eq_MCC}, evaluates the network's performance when dealing with sample sizes which widely differ in range, see \citep{matthews} and \citep{Boughorbel2017}.  Here, these samples are TP, FN, FP, TN.
\begin{equation}
\label{eq_MCC}
\text{ MCC} = \frac{\text{TP}\times \text{TN} - \text{FP}\times \text{FN}}{\sqrt{(\text{TP}+\text{FP})\times (\text{TP} + \text{FN} )\times (\text{TN} + \text{FP})\times (\text{TN} + \text{FN})}}
\end{equation}
A value of $-1$ would indicate that classification and data are totally anti-correlated, a value of $0$ would be a total random classification with respect to the data, while a value of $1$ would indicate a total correlation between the classification and the data. 
The MCC for the RNN is $\approx 0.42$. Comparing this to the accuracy it becomes clear that the MCC is a more robust way to evaluate the RNN's performance than the accuracy by itself.
In this context also the F1 score, see Eq.\ref{eq_F1} can be used to evaluate the accuracy of the RNN due to the model being binary, see \citep{doi:10.1002/asi.4630300621} and \citep{Powers2011}. At a value of 0, F1 indicates worst precision while a perfect precision is indicated at a value of 1. The F1 value of this RNN's capability to distinguish between RFI and non-RFI is $\approx 0.31$.
\begin{equation}
\label{eq_F1}
\text{ F1} = \frac{2\times \text{TP}}{2\times \text{TP} + \text{FP}+ \text{FN}}
\end{equation}

\begin{center}

\begin{table}[t]%
\centering
\caption{The results of the evaluation of the confusion matrix are listed here. The different parameters are calculated in accordance to \citep{matthews}, \citep{doi:10.1002/asi.4630300621},  \citep{Boughorbel2017}, \citep{Fawcett2006} and \citep{Powers2011}.\label{tab1}}%
\tabcolsep=0pt%
\begin{tabular*}{20pc}{@{\extracolsep\fill}lcc@{\extracolsep\fill}}
\toprule
\textbf{Parameter} & \textbf{Value}  \\

accuracy$^{\text{a)}}$ & 0.9792 \\
\midrule
positive predictive value$^{\text{b)}}$& 0.1800\\
false discovery rate$^{\text{c)}}$ & 0.8200\\
\midrule
negative predictive value$^{\text{d)}}$& 0.9999\\
false omission rate$^{\text{e)}}$ & $6.5\times10^{-6}$\\
\midrule
true positive rate$^{\text{f)}}$ & 0.9986\\
false negative rate$^{\text{g)}}$ & 0.0014\\
\midrule
true negative rate$^{\text{h)}}$ & 0.9791\\
false positive rate$^{\text{i)}}$ & 0.0209\\
\midrule
Mattew's correlation coefficient$^{\text{j)}}$ & 0.4195\\
F1 score$^{\text{k)}}$ & 0.3051\\
\bottomrule
\end{tabular*}
\begin{tablenotes}
a)  Eq.\ref{eq_acc}, b) Eq.\ref{eq_PPV}, c)  Eq.\ref{eq_FDR}, d) Eq.\ref{eq_NPV}, e) Eq.\ref{eq_FOR}, f)   Eq.\ref{eq_TPR}, g)  Eq.\ref{eq_FNR} ,h) Eq.\ref{eq_TNR}, i) Eq.\ref{eq_FPR}, j) Eq.\ref{eq_MCC}, k) Eq.\ref{eq_F1}
\end{tablenotes}
\end{table}
\end{center}

\section{ Conclusion and Outlook}
In Sec.\ref{perf}, we show that the RNN reaches an accuracy of $\approx 98\%$ after sufficient training. However, this seemingly high accuracy
is due to the large number of data in the TN category, see Eq.\ref{confmatr}. When studying the PPV and the FDR, a weakness of the chosen
method becomes apparent which lowers its overall efficiency. A large amount of data (FP category, $415179$ data points), which are
actually non-RFI, are classified as being RFI resulting in a PPV of $\approx 18\,\%$. This also becomes evident when taking the MCC into account, which is $\approx 0.42$, meaning the classification is not random with respect to the data, but the correlation is not strong either. When calculating the F1 score, the overall precision of the network is $\approx 0.31$.  An improvement of the efficiency of the method can be expected from the 
following refinements:
\begin{itemize}
\item \textbf{Data usage:} In this study, we used only the amplitude information in the data.  However, the amplitude differences with respect
to the channles, baselines, and times steps should also be used, adding four more axes to the training data cube.  In addition, the phase (spatial)
information in the data could be further utilized.
\item \textbf{Model complexity:} The discrete amplitude-based model to distinguish RFI and non-RFI may be adjusted to cope with more complex signal shapes and strength patterns.
\item \textbf{Network architecture:} The network training could be extended to consider the time step, polarization and baseline sequences instead of
the channel sequence only. Thus, the amount of data in the FP category will be reduced. By also adding the information on the image-level, it is possible to combine the RNN with the advantages of a CNN which would give information of prominent features in an image, giving a hierarchy of dominant features like RFI. This would result in a change of the architecture into a recurrent convolutional neural network (RCNN)
\end{itemize}
The results of this study mark an encouraging milestone and path towards a highly dynamical RFI filter meeting the challenges of future
radio antenna arrays.

\section*{Acknowledgments}
This research was supported by the
Bayerisch-Tschechische Hochschulagentur (BTHA)
under grant number BTHA-AP-2018-18

\subsection*{Author contributions}
All authors contributed to the scientific content, the writing and editing of the manuscript. The original analysis was done by PRB, TM, and JR.

\subsection*{Financial disclosure}
None reported.

\subsection*{Conflict of interest}
The authors declare no potential conflict of interests.

\bibliography{Wiley-ASNA}

\end{document}